\documentstyle[12pt]{article}
\def\be{\begin{equation}}
\def\ee{\end{equation}}
\def\bea{\begin{eqnarray}}
\def\eea{\end{eqnarray}}
\def\nn{\nonumber}
\def\bi{\bibitem}
\def\d{{\rm d}}
\def\e{{\rm e}}
\def\ii{{\rm i}}
\def\k{\kappa}
\def\ve{\varepsilon}
\def\a{\alpha}
\def\b{\beta}
\def\p{\partial}
\def\s{\sigma}
\def\o{\omega}
\def\sq{\sqrt{-g}}
\def\I{\int{\rm d}^2\xi}
\begin{document}
\begin{titlepage}
\title{Quantum Fluctuations and Curvature Singularities in
Jackiw-Teitelboim
gravity}
\author{Ctirad Klim\v c\'\i k\\ Theory Division of the Nuclear
Centre,
Charles University,\\ V Hole\v sovi\v ck\' ach 2, Prague 8, Czech
Republic}
\date{March, 1993}
\maketitle
\begin{abstract}
The Jackiw-Teitelboim gravity with the matter degrees of freedom
is considered. The classical model is exactly solvable and its
solutions
describe non-trivial gravitational scattering of matter wave-packets.
For huge amount of the solutions the scattering space-times are free
of
curvature singularities. However, the quantum corrections to the
field
equations inevitably cause the formation of (thunderbolt) curvature
singularities, vanishing only in the limit $\hbar\to 0$. The
singularities
cut the space-time and disallow propagation to the future.The model
is
inspired by the dimensional reduction of 4-d pure Einstein gravity,
restricted to the space-times with two commuting space-like Killing
vectors. The matter degrees of freedom also stem from the 4-d ansatz.
The measures for the continual integrations are judiciously chosen
and one loop contributions (including the graviton and the dilaton
ones)
are evaluated.
For the number of the matter fields $N=24$ we obtain even the exact
effective
action, applying the DDK-procedure.
The effective action is
nonlocal, but the semiclassical equations can be solved
by using some
theory of the Hankel transformations.

\end{abstract}
\end{titlepage}
\section{Introduction}

The Einstein gravity is undoubtedly beautiful and physically relevant
theoretical construction which, at the same time, brought many novel
mathematical structures into theoretical physics.
  The notions like the black-hole, the horizon,
the curvature singularities or the gravitational collapse enriched
our
conceptual world, but they also posed new
challenging problems, yet to be understood. For instance, what is the
fate
of matter in the last part of the gravitational collapse? Following
the
overwhelming success of the quantum theory in describing the world of
subatomic distances, physicists feel that an appropriate theory of
quantum gravity has to provide the correct solutions to the problems
and
improve our understanding of the phenomena, occuring in very strong
gravitational fields near the curvature singularities. Unfortunately,
such
generally accepted and technically applicable quantum theory of
gravity does
not exist yet. The considerable breakthrough was reached due to
string theory,
where the consistent perturbative $S$ matrix for scattering of
gravitons
and excitations of other fields can be obtained \cite{Gre}. However,
the
dynamics of a gravitational collapse or the status of quantum black
holes
remain nonunderstood. Though there is a common belief that quantum
effects
should smear the singular behaviour of the classical theory, there is
no
sufficiently established quantitative evidence for such a conjecture.

     The string boom had also an indirect, but very important
influence
on the subject of quantum gravity. The models of $1+1$ dimensional
theories
of gravity coupled to a dilaton field $\phi$ have arisen in string
theory
\cite{Witt,Man} . These models possess
black hole solutions and they motivated
Callan, Giddings, Harvey and Strominger \cite {CGHS} to investigate
the
``$CGHS$" model of 2-d dilaton gravity coupled to conformal matter.
     The model is of interest as a ``toy" model of quantum gravity in
two
dimensions which contains gravitational collapse, black holes, cosmic
censorship and Hawking radiation. Moreover, the model is very similar
to that
obtained by the dimensional reduction of the spherically symmetric
gravitational system in $3+1$ dimensions, hence, one may expect the
relevance
of the $1+1$ results to the $3+1$ physics. Recently, many authors
have been
investigating the quantum dynamics of the black holes by using the
$CGHS$
model \cite{BDDO}-\cite{Stro}.
The issues of particular interest are the backreaction of
the Hawking radiation on the metric and the endpoint of the black
hole
evaporation. The problem is far simpler than the original $3+1$
dimensional one and powerful methods of conformal field theory in two
dimensions can be used.

The $CGHS$ model and its variants \cite{BilCal,DeA2} and also other
2-d dilaton gravities have been
studied in the literature \cite{Tri}, particularly in the context of
the
noncritical string theory \cite{Cham}. The models can be typically
obtained
by the dimensional reduction of the higher-dimensional pure metric
gravities.
This fact suggests the following $CGHS$-like scenario for
``addressing" the
four-dimensional quantum problems: one finds the corresponding
dilaton gravity
model in 1+1 dimensions and attempts to quantize it. Though the 1+1
quantum
theory may still be complicated enough to prevent exact solvability
(like
$CGHS$ is), usually it is far simpler than its 4-d counterpart. In
this
contribution, we adopt the scenario and address the quantum dynamics
of the
colliding gravitational waves. The fact, that the nonlinear character
of the
Einstein equations results in the formation of curvature
singularities after
collisions of gravitational waves, is known only since seventies
\cite{Sz,Kh}
and, perhaps, it is less familiar to nonspecialists than the fact
that a
black hole is formed as the consequence of gravitational collapse.
However,
the colliding-waves problem keeps attracting many relativists
\cite{NH}-\cite{Grif}
without an interruption, since the discoveries of the first
colliding-waves
space-times by Szekeres \cite{Sz} and Khan and Penrose \cite{Kh}.
The problem of main interest for us will constitute in the following:
What is
the quantum status of the those
scattering space-times? As we have mentioned above,
one usually expects that the curvature singularities should be
smeared by the
effects of quantum fluctuations. Our quantitative analysis will show,
however,
the surprising result: the quantum curvature singularities are even
worse than
the classical ones and even classicaly nonsingular spacetimes are
destabilized
by quantum curvature singularities.

  Apart from the physical questions which our analysis will try to
answer,
the model to be considered in this paper is of interest also for some
more
theoretical
reasons. Indeed, while in $CGHS$ and related theories \cite{Tri,Cham}
the matter degrees of freedom are added by hand {\it after} the
dimensional
reduction, in our model the matter degrees of freedom also stem from
the
four-dimensional theory. This fact should even increase the relevance
of our
results for the 4-d case. There is another pleasant thing, namely,
not only
the matter loops, but also one loop dilaton and graviton
contributions can be
evaluated, yielding the one loop effective action. We need not to
perform
$1/N$ expansion, but we can take $\hbar$ as the natural parameter of
expansion. Moreover, for the critical number of matter fields (N=24)
our
result will be nonperturbative and exact. But the good news is not
exhausted
by that, it turns out, moreover, that the semiclassical equations can
be
solved and the behaviour of curvature singularities is under control.

The plan of the paper is as follows. In section 2 we introduce the
2-d
matter-dilaton model motivated
by the dimensional reduction of the $3+1$ dimensional system
with two commuting space-like Killing vectors. Then we find the
classical
equations of motion in the conformal gauge. The dilaton field turns
out to
satisfy the standard d'Alambert wave equation, hence, we introduce
a sort of
the ``light cone" gauge. In this gauge the matter fields obey the
Gowdy
cylindrical wave equation \cite{Gow}, the general solution of which
can be
given by the decomposition into the Fourier-Bessel and
Fourier-Neumann modes.
 The corresponding metric we find explicitly by integrating the
remaining
equations. We show that the Neumann modes cause the formation of the
(classical) curvature
singularities which close the space-time to the future, while the
appropriate
superpositions of the Bessel modes describe the collisions of the
wave packets
travelling against each other with the velocity of light. The
corresponding
spacetimes are everywhere regular with the out region in which the
scattered
wave packets travel to the opposite space infinities. In section 3 we
discuss
the quantization of the model. We choose the standard Polyakov
measure for the
functional integration over the metrics and the reparametrization
invariant
measures for the dilaton and the matter fields integration. Then we
compute
the one loop effective action . The effective action is nonlocal even
in the
conformal gauge, due to the presence of the direct matter-dilaton
coupling in
the action. The one loop effective field equations are localized by
going
to the dilaton ``light-cone" gauge.
 The renormalization requires a
purely dilatonic counterterm, the contribution of
which makes finite one infinite constant in the
semiclassical field equations. The actual computation requires the
knowledge
of the functional derivatives of the determinant of the (Gowdy) wave
operator
with respect to the dilaton and metric. They are evaluated by using
the heat
kernel regularization and some theory of the Hankel transformations
in the Appendix. In section
4 we solve the semiclassical field equations. We perform the detailed
analysis
of the scalar curvature of the space-times, obtained by solving the
semiclassical equations. We show that the contribution of the quantum
fluctuations to the effective action inevitably generates the
curvature
singularities in the semiclassical spacetimes. These singularities
may
disappear only in the limit $\hbar\to 0$, thus indicating, that the
classical
regular scattering space-times are in fact unstable from the quantum
point
of view. We end up with short conclusions and outlook.

\section{The model and its classical dynamics}

\subsection{The dimensional reduction}

The form of the four-dimensional metric describing the collisions of
collinear
gravitational waves is given by \cite{Grif}

\be {\d}s^2 = -2\phi^{-{1\over 2}}{\e}^{\mu}\d u\d v +
\phi(\e^{\sqrt{2\kappa}
Q}\d x^2 +\e^{-\sqrt{2\kappa}Q}\d y^2),\label{ansatz} \ee
where the metric functions $\mu,\phi$ and $Q$ are invariant on the
$(x,y)$
plane of symmetry. The 4-d vacuum Einstein equations for the metric
(\ref{ansatz})
consist of the constraints

\bea -\phi_{uu}+\mu_u\phi_u=\kappa\phi Q_u^2 \label{con1}\\
-\phi_{vv}+\mu_v\phi_v=\kappa\phi Q_v^2\label{con2} \eea
and the evolution equations

\be \phi_{uv}=0 \label{2.1.3a} \ee
\be (\phi Q_u)_v +(\phi Q_v)_u =0 \label{2.1.3b} \ee
\be -\mu_{uv}=\kappa Q_u Q_v \label{2.1.3c}.\ee

It is not difficult to demonstrate, that the same set of the
constraints
and the evolution equations follows from the following 2-d action

\be S={1\over 2\kappa}\int \d^2\xi\sq \phi (R-\k g^{\a\b}\p_{\a} Q
\p_{\b} Q), \label{2.1.4}\ee
after fixing the conformal gauge

\be \d s^2=-2\e^{\mu}\d u\d v .\label{2.1.5}\ee
The action (\ref{2.1.4}) can be interpreted as the Jackiw-Teitelboim
gravity
\cite{Jack}
where the cosmological constant is replaced with the kinetic term of
the
matter.
 The matter is coupled to the dilaton and possesses all
dynamical degrees of freedom of the theory.

\subsection{The solutions of the field equations}

In what follows, we shall consider the model (\ref{2.1.4}) with an
arbitrary
number of matter fields. The classical dynamics does not change
dramatically, but the properties of the quantum theory will depend on
that
number. The action reads

\be S={1\over 2\kappa}\int \d^2\xi\sqrt{-g}\phi(R-\kappa
g^{\a\b}\p_{\a}Q^j
\p_{\b}Q^j) \label{2.2.1}
\ee
and the constraints and the evolution equations in the conformal
gauge get
obviously modified

\bea -\phi_{uu}+\mu_u\phi_u=\kappa\phi Q_u^j Q_u^j\label{2.2.2a}\\
-\phi_{vv}+\mu_v\phi_v=\kappa\phi Q_v^j Q_v^j\label{2.2.2b} \eea
\be \phi_{uv}=0 \label{2.2.3a} \ee
\be (\phi Q_u^j)_v +(\phi Q_v^j)_u =0 \label{2.2.3b} \ee
\be -\mu_{uv}=\kappa Q_u^j Q_v^j \label{2.2.3c}.\ee
The general solution of (\ref{2.2.3a}) reads

\be \phi = f(u) +g(v) \label{2.2.4} \ee

If $\phi_v =0$ (or $\phi_u =0$), then from (\ref{2.2.2b})
(or (\ref{2.2.2a})) it follows that
$Q_v^j=0$ ($Q_u^j=0$) and from (\ref{2.2.3c}) $\mu_{uv}=0$. Since the
scalar
curvature $R$ is given by
\be R=2\e^{-\mu}\mu_{uv} \label{2.2.5} \ee
and in two dimension the curvature tensor reads

\be R_{\a\b\gamma\delta}={1\over
2}(g_{\a\gamma}g_{\b\delta}-g_{\a\delta}
g_{\b\gamma})R \label{2.2.6} \ee
we may conclude that arbitrary functions $\phi(u)$ and $Q^j(u)$
(or $\phi(v)$ and $Q^j(v)$) are solutions of the field equations and
the
corresponding space-time is flat. Such solutions obviously describe
the matter
excitations propagating in one direction with the velocity of light.

If neither $\phi_v=0$ nor $\phi_u =0$ we can (at least locally)
perform the
conformal transformation

\be U=\sqrt{2} f(u), V=\sqrt{2} g(v) \label{2.2.7} \ee
Such a transformation is obviously the symmetry transformation of the
set
of the field equations (\ref{2.2.2a}-\ref{2.2.3c}),
hence we may fix this residual symmetry
by the claim

\be \phi={U+V\over \sqrt{2}}\equiv t\label{2.2.8}\ee
We may call this gauge fixing the ``dilaton" gauge. In the gauge
the field equations become

\bea \mu_U=\k (U+V)Q_U^j Q_U^j \label{2.2.9a} \\
\mu_V=\k(U+V)Q_V^j Q_V^j \label{2.2.9b} \eea

\be Q_{tt}^j+{1\over t}Q_t^j-Q_{\s\s}^j=0 \label{2.2.10} \ee
\be -\mu_{UV}=\k Q_U^j Q_V^j \label{2.2.10c} \ee
where

\be \s\equiv {V-U\over \sqrt{2}} \nn \ee
We observe that the linear equation for the matter fields $Q^j$
does not contain
the metric function $\mu$. We may call this equation by the name of
Gowdy,
who studied cosmological models with plane symmetry \cite{Gow},
governed locally by
(\ref{2.2.10}). The general solution of the Gowdy equation (which
tend to zero
at the spatial infinities) is given by the following mode expansion
\cite{Grif,FeIb}.

\be Q^j(t,\s) =\int_0^{\infty} \d \o {\rm Re}\big[
A^j(\o)J_0(\o t)\e^{-\ii\o\s}
+ B^j(\o)N_0(\o t)\e^{-\ii\o\s}\big], \label{2.2.12} \ee
where $A^j$ and $B^j$ are (complex) distributions ensuring the proper
behaviour at the space infinity and $J_0$ and $N_0$ are the Bessel
and Neumann functions of zero order, respectively.
 This mode expansion can be
easily found by using the Fourier tranformation in the variable
$\s$ in (\ref {2.2.10}). The resulting ordinary differential equation
in the
variable $t$ is then the Bessel equation. We should note, at this
place,
that in higher dimensions some additional (so called ``solitonic")
terms are considered at the r.h.s.~of
(\ref{2.2.12}). They do not vanish at the space infinity and in the
limit
of the weak gravitational coupling $\k\to 0$ those solutions diverge
and
do not approach the non-interacting matter solutions \cite{kliko}. We
shall
not consider this ``solitonic" sector in this paper and prescribe the
boundary conditions, mentioned above.

It remains to solve the equations (\ref{2.2.9a}),(\ref{2.2.9b})
and (\ref{2.2.10c}),
which determine
the metric function $\mu$. Combining (\ref{2.2.9a}) with
(\ref{2.2.9b}),
we obtain

\be \mu_{\s}=2\k t Q_t^j Q_{\s}^j \label{2.2.13a} \ee

\be \mu_t=\k t(Q_t^j Q_t^j + Q_{\s}^j Q_{\s}^j) \label {2.2.13b} \ee
     We use the fact \cite{Gray} that for $F$ and $G$, fulfilling the
Bessel
equations

\bea x^2{\d^2 F\over \d x^2} + x{\d F\over \d x} +(\lambda^2 x^2
- n^2)F=0
\nn\\x^2{\d^2 G\over \d x^2} +x{\d G\over \d x}+(\nu^2 x^2-n^2)G=0
\nn \eea
it holds

\be \int_a^b \d x (\lambda^2-\nu^2)x F~G = \Big[x(F{\d G\over \d x}-
G{\d F\over \d x})\Big]_a^b
\label{2.2.14}\ee
This formula enables us to integrate the products of the Bessel
functions.
The result of the integration gives the explicit form of the metric
function\footnote{It appears, that this result is new even from the
point of
view of 4-d theory of colliding waves.} $\mu$

\bea \mu =\k\int_0^{\infty}\d\o_1 \d\o_2 \o_1 \o_2~ t~{\rm
Re}\Big[{-1\over
\o_1 +\o_2}G_1^j(\o_1 t)G_0^j(\o_2 t)\e^{-\ii(\o_1 +\o_2)\s} \nn\\+
{1\over 2}{1\over
\o_1 -\o_2}\big(G_1^{j*}(\o_1 t)G_0^j(\o_2 t)\e^{\ii(\o_1-\o_2)\s}-
G_1^{j*}(\o_2 t)G_0^j(\o_1 t)\e^{-\ii(\o_1-\o_2)\s}\big)\Big],
\label{2.2.15} \eea
where
\be G_{0(1)}^j(\o t)=
A^j(\o)J_{0(1)}(\o t)+B^j(\o)N_{0(1)}(\o t) \label{2.2.16a}\ee
We note, that the classical equations (\ref{2.2.3a}-\ref{2.2.3c})
turn out
to be ``iteratively" linear. Indeed, solving the linear
Eq.(\ref{2.2.3a})
and inserting its solution $\phi$ into Eq.(\ref{2.2.3b}), we get
again the
linear equation. After solving it, we insert $Q^j$ into
Eq.(\ref{2.2.3c})
and get the linear equation for $\mu$. Such a structure of the
equations
gives the classical integrability and will be also important later
for the
quantization.
\subsection{Curvature singularities and the global structure}

We start our analysis of the curvature singularities with the formula
for the
scalar curvature. Following from Eqs. (\ref{2.2.5}) and
(\ref{2.2.10c}),
we have

\be R=\k\e^{-\mu}(-Q_t^j Q_t^j + Q_{\s}^j Q_{\s}^j).\label{2.2.16}\ee
Near $t \to 0^+$ we have

\be J_0(t)\sim 1-{t^2\over 4}+\cdots \label{2.2.17a}
\ee
\be N_0(t)\sim (1-{t^2\over 4})\ln{t} +\cdots \label{2.2.17b} \ee
hence,

\be Q_t^j\sim {1\over t}(\int_0^{\infty}\d\o \o {\rm
Re}[B^j(\o)\e^{-\ii\o\s}])
+{\rm bounded}\equiv {E^j\over t}+{\rm bounded},\label{2.2.18a}\ee
\be Q_{\s}^j\sim\ln{t}~(\int_0^{\infty}\d\o \o {\rm Re}[-\ii B^j(\o)
\e^{-\ii\o\s}])+{\rm bounded}\equiv H^j\ln{t} +{\rm
bounded},\label{2.2.18b}\ee
\be \mu\sim\k\ln{t} ~E^j E^j +{\rm bounded}.\label{2.2.18c}\ee
Inserting Eqs.(\ref{2.2.18a}),(\ref{2.2.18b}) and (\ref{2.2.18c})
 into Eq.(\ref{2.2.16}),
we have
\be R\sim t^{-\k E^j E^j}[-{E^j E^j\over t^2}+H^j H^j
(\ln{t})^2+\cdots].
\label{2.2.19}\ee
We conclude, that the regularity of the (classical) space-times
requires both
$E^j$ and $H^j$ to be equal zero, or, equivalently,

\be B^j(\o)=0.\label{2.2.20}\ee

Consider now (regular) space-times, given by

\be A^j(\o,\o_0,\rho)=\vert a_j\vert \e^{\ii\phi_j}\sqrt{{\o\over
4\rho}}
\e^{-{(\o-\o_0)^2\over 2\rho}},~ B^j(\o)=O,\label{2.2.23}\ee
where $\phi_j$ is real and $\rho$ and $\o_0$ are real positive
parameters.
Note that for $\rho\to 0$

\be A^j(\o,\o_0,\rho)\to\vert a_j\vert\e{\ii\phi_j}{\o_0\pi\over
\sqrt{2}}\delta(\o-\o_0).\ee
{}From Eq.(\ref{2.2.12}) for $B^j=0$ we obtain
\be Q^j=\int_0^{\infty}\d\o {\rm Re}\Big[\vert a_j\vert
\e^{\ii\phi_j}
\sqrt{{\o\over 4\rho}}\e^{-{(\o-\o_0)^2\over 2\rho}}
J_0(\o t)\e^{-\ii\o\s}\Big].\label{2.2.25}\ee
{}From the well-known formula for the asymptotic behaviour of the
Bessel
functions for $t\to\pm\infty$ \cite{Gray}
\be J_0(\o t)=\sqrt{{2\over \pi\o\vert t\vert}}\cos{(\o t \mp
{\pi\over 4})}
+\cdots,\label{2.2.26}\ee
we have for $t\to\pm\infty$
\bea Q^j &=&{\vert a_j\vert\over \sqrt{\vert
t\vert}}\Big\{\e^{-{\rho\over 2}
(t-\s)^2}\cos{[\o_0(t-\s)+\phi_j\mp{\pi\over 4}]}\nn\\
&+&\e^{-{\rho\over 2}(t+\s)^2}
\cos{[\o_0(t+\s)-\phi_j\mp{\pi\over 4}]}\Big\}.\label{2.2.27}\eea
Now the physical interpretation of this solution is obvious. At
$t\to -\infty$
we have two wave packets propagating against each other by the
velocity of light;
at $t\to\infty$ the two scattered packets propagate apart from each
other with
the gained phase shift, indicated in Eq.(\ref{2.2.27}). Because
$J_0(\o t)$
and its derivatives are bounded functions \cite{Gray}, we may use the
Riemann-Lebesgue lemma and from Eq.(\ref{2.2.25}) conclude that for
$t={\rm const}$ and $\s\to\pm\infty$,
$Q^j$ and all its derivatives with respect
to $t$ and $\s$ vanish. Hence, by using the constraints
(\ref{2.2.13a}) and
(\ref{2.2.13b}) and the formula (\ref{2.2.16}) for the scalar
curvature, we
observe that the space-time is flat in this limit. For the cases
$\s={\rm
const}$, $t\to\pm\infty$; $t+\s$=const, $t-\s\to\pm\infty$
 and $t-\s$=const, $t+\s\to\pm\infty$, we use
the asymptotic expression (\ref{2.2.27}), the constraints
(\ref{2.2.13a}),
(\ref{2.2.13b}) and the formula (\ref{2.2.16}) to arrive at
 the same conclusion.
Therefore, for the ``wave-packet" choice (\ref{2.2.23}) the
corresponding
space-time is asymptotically flat (see Figure 1),
 it has the same topology as the
two-dimensional Minkowski space-time and is free from curvature
singularities.
We shall not need the explicit form of the metric, which,
nevertheless,
can be obtained by performing the integral (\ref{2.2.15}) with the
choice
(\ref{2.2.23}).
\setlength{\unitlength}{1mm}
\begin{figure}
\begin{center}
\begin{picture}(90,70)

\put(45,0){\vector(0,1){70}}
\put(45,35){\oval(45,33)}
\put(0,35){\vector(1,0){90}}
\put(45,42){\makebox(0,0){~~~~$R\neq0,$~~~$R$ bounded}}
\put(70,60){\makebox(0,0){$R\to 0$}}
\put(48,65){\makebox(0,0){$t$}}
\put(85,32){\makebox(0,0){$\s$}}
\end{picture}
\caption[]{The scalar curvature of the regular classical
space-times.}
\end{center}
\end{figure}
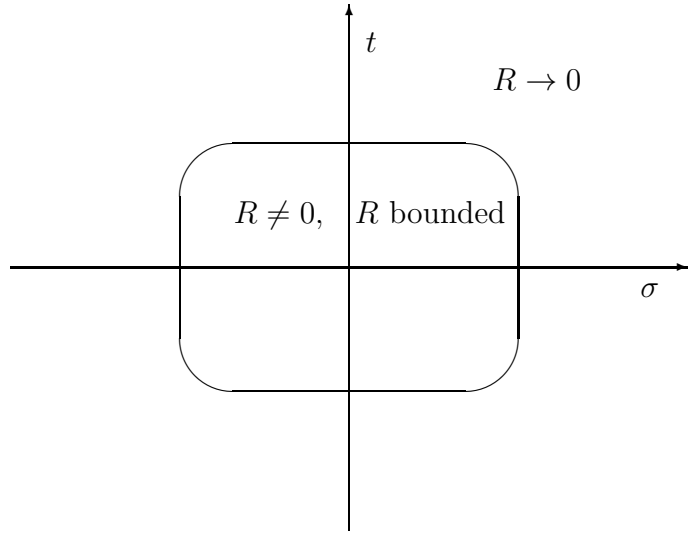
We should end up the classical
analysis with some important comments.

First of all, it does not seem unexpected that for the collisions of
the
localized wave packets travelling with the velocity of light, the
space-time is asymptotically flat in the space-like and time-like
directions.
 What looks more surprising is the fact
 that the same is true for the null infinities.
The reason is simple: in two dimensions a {\it single} propagating
wave
does not curve the space-time (cf. the analysis after
(Eq.\ref{2.2.6})).
In higher dimensions this is not true \cite{Garr,kliko}, but in that
case
the curvature is given by the shape of the wave front in the {\it
transverse}
directions. Since there are no transverse directions in two
dimensions, our
result could be anticipated.

The second comment is closely related to the first one. It concerns
the
regularity of the initial data. In the higher dimensional case the
following problem was studied \cite{Hay,kliko}. If initially regular
gravitational waves interact, will a curvature singularity be formed?
The criterion for the regularity of the incoming data can be
naturally
formulated:
one requires the boundedness of the
amplitude of the wave,which itself is defined by means of the
components
of the Riemann tensor of the corresponding metric
in the so called parallelly propagated orthonormal
frame \cite{Hay,kliko}.
 However, in the two-dimensional case, incoming waves do not curve
the spacetime and this criterion of regularity fails.
 But certainly we should not consider all incoming waves
as regular (with bounded amplitude). One might require that the
scalar
fields $Q^j$
itself should be bounded, but, on the other hand, we could rescale
it by an arbitrary function of another scalar field, the dilaton
$\phi$,
and we would get the classically equivalent dynamical theory with the
different condition of incoming regularity. Fortunately, it appears
to be a
natural candidate for the amplitude of the wave.
The matter part of the action (\ref{2.2.1})
suggests the following inner product on the space of fields $Q$
(see also Eq.(\ref{3.1.4}))
\be (Q_1,Q_2)=\int \d^2\xi\sqrt{-g}\phi Q_1 Q_2.\label{2.2.28} \ee
Hence, our condition of the incoming regularity reads

\be \displaystyle\lim_{U\to\infty(V\to\infty)}\phi Q^j Q^j={\rm
finite}
\label{2.2.29}\ee
In the case of the wave-packets (\ref{2.2.23}) we get

\be \displaystyle\lim_{U\to\pm\infty}\phi Q^j Q^j= \sum_j \vert
a_j\vert^2
\e^{-2\rho V^2} {\rm cos}^2[\sqrt{2}\o_2 V-\phi_j\mp{\pi\over 4}]=
{\rm finite}\ee
and similarly for $V\to\pm\infty$.

\section{The quantization}
\subsection{The functional measures and the effective action}

Define the generating functional of the model $W[J_g,J_{\phi},J_j]$
by
\bea& &W[J_g,J_{\phi},J_j]\equiv\int{D_{g}g_{\a\b}D_{g}\phi
D_{g,\phi}Q^j\over
{\rm Vol}({\textstyle Diff})}\label{3.1.1}\\
& &\exp{\Big\{{\ii\over \hbar }\int\d^2\xi\big[\sqrt{-g}{\phi\over
2\k}(R-\k g^{\a\b}\p_{\a}Q^j\p_{\b}Q^j)+\sqrt{-g}J_gR +J_{\phi}\phi
+J_j Q_j\big]\Big\}}\nn \eea
where $J_g$ is a scalar source, $J_{\phi}$ and $J_j$ are scalar
densities
and ${\rm Vol}({\textstyle Diff})$ is the volume of the group of
diffeomorphisms.
We define the functional measures by the following norms
\be \vert\vert \delta g\vert\vert^2=\int\d^2\xi\sq
g^{\a\gamma}g^{\b\delta}
\delta g_{\a\b}\delta g_{\gamma\delta},\label{3.1.2}\ee
\be \Vert\delta\phi \Vert^2 =\I\sq \delta \phi^2,\label{3.1.3}\ee
\be\Vert\delta Q^j \Vert^2 =\I\sq ~\phi~\delta Q^j
Q^j.\label{3.1.4}\ee
Eq.(\ref{3.1.2}) defines the usual de Witt-Polyakov norm \cite{Pol}
and
Eq.(\ref{3.1.3}) gives the standard reparametrization invariant
measure for a
scalar field. The norm (\ref{3.1.4}) is given by the form of the
matter
part of the classical action, much in the same way as the norm for
the
quantization
of the
standard non-linear $\sigma$-model with the coordinates
$X^A$ of the target and a metric $H_{AB}(X)$, i.e. \cite{Frie}
\be \Vert \delta X^j \Vert^2=\I\sq H_{AB}(X(\xi))
\delta X^A(\xi)\delta X^B(\xi).\label{3.1.5}\ee
We return to Eq.(\ref{3.1.1}) and we fix the conformal gauge
\be \d s^2=-2\e^{\mu}\d u\d v.\label{3.1.6}\ee
By using the standard Faddeev-Popov procedure, we obtain
\be W[J]=\int D_{\mu}\mu D_{\mu}\phi D_{\mu,\phi}Q^j\exp{\big\{\ii
{26\over 48\pi}\I{1\over 2}\mu\p^2 \mu\big\}}\nn\ee
\be \times\exp{\Big\{{\ii\over \hbar}
\I\big[{\phi\over 2\k}(-\p^2\mu -\k\p Q^j \p Q^j)
-J_g\p^2\mu +J_{\phi}\phi +J_j Q_j\big]\Big\}},\label{3.1.7}\ee
where $\p^2$ is the Minkowski d'Alambertian and $\p Q^j \p Q^j$
means the Minkowski metric scalar product. The Weyl anomaly
term comes from the Faddeev-Popov determinant and the measure
$D_{\mu}\mu$
is given by the following norm
\be \Vert \delta\mu \Vert^2=\I\e^{\mu}\delta\mu^2.\label{3.1.8}\ee
We standardly
suppose that the exponential term in the Weyl anomaly is eventually
(after including all contributions)
cancelled by tuning of the 2-d cosmological constant counter-term.

Define now the effective action $\Gamma$ by the following
prescription

\be {h\over \ii}\ln{W[J]}\equiv Z[J]\equiv \Gamma[\mu_c,\phi_c,Q_c^j]
-J_g\p^2\mu_c +J_{\phi}\phi_c +J_j Q_c^j,\label{3.1.9}\ee
where
\be \p^2\mu_c\equiv -{\delta Z\over \delta J_g},~~~\phi_c\equiv
{\delta Z\over \delta J_{\phi}},~~~Q_c^j\equiv {\delta Z\over \delta
J_j}.
\label{3.1.10}\ee
We wish to compute the one loop effective action $\Gamma_1$. In order
to do
that, we have first to determine $Z_1$ (the generating functional for
the
connected Green functions) from (\ref{3.1.7}) and then to perform the
Legendre transformation (\ref{3.1.9}) and (\ref{3.1.10}). We note,
that the
dependences of the measures on the fields $\mu$ and $\phi$ are of the
order $O(\hbar)$ with respect to the classical action in the
exponent. Hence,
the loop diagrams
with the vertices coming from the measures will be of the order
$O(\hbar^2)$ and may be neglected in the one loop approximation.
Therefore,
we may write

\bea W_{semicl.}[J]=\int D_{\mu_J}\mu D_{\mu_J}\phi
D_{\mu_J,\phi_J}Q^j
\exp{\big\{\ii{26\over 96\pi}\I~ \mu_J\p^2\mu_J\big\}} \nn\\
\exp{\Big\{{\ii\over \hbar}\I \big[{\phi\over 2\k}(-\p^2\mu-\k\p Q^j
\p Q^j)
-J_g\p^2\mu +J_\phi \phi +J_j Q^j\big]\Big\}},\label{3.1.11}\eea
where $\mu_J$ and $\phi_J$ are the saddle point values of the
exponent, given by the equations

\be -\p^2\mu_J-\k\p Q^j \p Q^j + 2\k J_{\phi}=0,\label{3.1.12}\ee
\be \phi_J+2\k J_g=0,\label{3.1.13}\ee
\be \p(\phi_J \p Q^j_J)+J_j=0.\label{3.1.14}\ee
We observe that in the one loop approximation, we can consider the
measures to be independent on the field integration variables (but,
of course, dependent on the Schwinger currents). Now we evaluate the
integral
(\ref{3.1.11}). In the (second) exponent, there stands the quadratic
form
in the variables $\phi$ and $\mu$. Moreover, the norms defining the
measures
has the same $\mu_J$-dependence., i.e.

\be \Vert \delta \mu \Vert^2=\I\e^{\mu_J}\delta\mu^2; \Vert \delta
\phi
\Vert^2=\I\e^{\mu_J}\delta \phi^2.\label{3.1.15}\ee
Therefore, we can easily perform the Gaussian integration over $\mu$
and
$\phi$ with the result
\bea W_{semicl.}[J]=\int D_{\mu_J,\phi_J}Q^j\exp{\big\{\ii{24\over
96\pi}\I~
\mu_J\p^2 \mu_J\big\}}\nn\\
\exp{\Big\{{\ii\over \hbar}\I(-2\k J_g J_{\phi}+\k J_g\p Q^j \p Q^j +
J_j Q^j)\Big\}}.\label{3.1.16} \eea
The integration over $Q^j$ is again Gaussian, hence we obtain the
closed
expression for the semiclassical generating functional

\bea W_{semicl.}[J_g,J_\phi,J_j]=
{\rm det}^{-N/2}\big[-\ii{\k\over \hbar}\e^{-\mu_J}{1\over 2\k J_g}
\p(J_g \p)\big]\nn\\
\exp{\Big\{\ii{24\over 96\pi}\I\mu_J
\p^2 \mu_J +{ \ii\over \hbar }\I(-2\k J_g J_{\phi}+J_j{1\over
4\k\p(J_g \p)}
J_j)\Big\}},\label{3.1.17}\eea
where we used the definition (\ref{3.1.4}) of the measure
$D_{\mu,\phi}Q^j$
and Eq.(\ref{3.1.13}). The Legendre transformation (\ref{3.1.9}) and
(\ref{3.1.10}) can be easily performed and we obtain the following
expression
for the one loop effective action

\bea \Gamma_1(\mu_c,\phi_c,Q^j_c)=\I{\phi_c\over 2\k}(-\p^2 \mu_c-\k
\p Q_c^j \p Q_c^j)+\hbar {24\over 96\pi}\I \mu_c \p^2 \mu_c \nn\\
+\ii\hbar {N\over 2}{\rm lndet}\big[-{\ii\over
2\hbar}\e^{-\mu_c}{1\over
\phi_c}\p(\phi_c \p)\big]+O(\hbar^2).\label{3.1.18}\eea
We recognize the classical action and two quantum corrections. The
first
one is the Weyl anomaly, while the second one is the nonlocal term
depending on $\mu_c$ and $\phi_c$. The expression (\ref{3.1.18}) can
also
be written in the manifestly covariant way, i.e.
\bea \Gamma_1(g_{c,\a\b},\phi_c,Q_c^j)=\I\sqrt{-g_c}~{\phi_c\over
2\k}
(R_c-\k g_c^{\a\b}\p_{\a} Q_c^j \p_{\b} Q_c^j)\nn\\
+\hbar{24\over 96\pi}\I\sqrt{-g_c} R_c({1\over \sqrt{-g_c}}\p_{\a}
\sqrt{-g_c} g_c^{\a\b} \p_{\b})^{-1}R_c \nn\\+
\ii\hbar {N\over 2}{\rm lndet}
\big[-{\ii\over 2\hbar}{1\over
\sqrt{-g_c}\phi_c}\p_{\a}(\sqrt{-g_c}\phi_c
g_c^{\a\b}\p_{\b})\big]+O(\hbar^2).\label{3.1.19}\eea

\subsection{The case $N=24$}

For $N=24$, the quadratic term of the Weyl anomaly vanishes.
The dilaton gravities usually get simplified and more precise results
can be obtained in that case \cite{Verl}. This happens also in our
model.
We show that the semiclassical effective action $\Gamma_1$
(\ref{3.1.18})
is the exact quantum effective action of the theory for $N=24$. We
use the
DDK-approach \cite{DDK1,DDK2}
 to establish this result. The dependences of the
functional measures on the field $\mu$ read
\be D_{\mu}\mu=(D_0 \mu)\exp{[-{\ii\over 48\pi}\I{1\over 2}\mu \p^2
\mu]},
\label{3.4.1}\ee
\be D_{\mu}\phi=(D_0 \phi)\exp{[-{\ii\over 48\pi}\I{1\over 2}\mu
\p^2 \mu]},
\label{3.4.2}\ee
\bea D_{\mu,\phi}Q^j&=&(D_{0,\phi}Q^j)\label{3.4.3}\\ & &\times
\exp{\big[-{\ii\over 48\pi}\I\{{1\over 2}\mu \p^2 \mu +{3\over 2}\mu
[2\p^2\ln{\vert \phi \vert}+(\p\ln{\vert \phi \vert})^2]\}\big]}.\nn
\eea
The formulas (\ref{3.4.1}) and (\ref{3.4.2}) are fairly standard
\cite{Mira,Kurz},
however, the relation (\ref{3.4.3}) deserves some comment. Indeed, it
can be
explicitly derived by computing the Jacobian, which relates both
measures,
with some regularization procedure. We shall use the heat kernel
 regularization and use the defining formula (\ref{3.1.4}) to write
\be D_{\mu,\phi}Q=D_{0,\phi}Q\sqrt{{\rm det}L},\label{3.4.24}\ee
where $L$ is the diagonal operator
\be L(\xi_1,\xi_2)=\e^{\mu(\xi_1)}\delta
(\xi_1,\xi_2).\label{3.4.25}\ee
Note that the $\delta$-function $\delta(\xi_1,\xi_2)$ is to be
understood
in the sence of the scalar product (\ref{3.1.4}) with $\mu=0$, i.e.
\be \delta(\xi_1,\xi_2)={1\over
\phi(\xi_1)}\delta(\xi_1-\xi_2).\label{3.4.26}
\ee
Clearly
\be \delta{\rm lndet}~L=\delta {\rm
Trln}~L=\I\phi(\xi)\delta(\xi,\xi)
\delta\mu(\xi),\label{3.4.28}\ee
where $\delta(\xi,\xi)$ is the meaningless quantity.As
in \cite{Mira,Kurz}, we replace it by the
heat kernel of the covariant Laplacian, but in our case
 with respect to the scalar product
(\ref{3.1.4}), i.e.
\bea \delta_{\ve}(\xi,\xi)={1\over \phi(\xi)}
<\xi\vert \exp{\Big\{-\ve\big[-{\ii\over \hbar}{1\over\sq \phi}
\p_{\a}(\sq\phi g^{\a\b}\p_{\b})\big]\Big\}}\vert \xi>
=\nn\\
={1\over \phi(\xi)}(-\ii)
\big\{{1\over 24\pi} \p^2 \mu +{1\over 16}
[2\p^2\ln{\vert \phi \vert}+(\p\ln{\vert \phi \vert})^2]\big\}
.\label{3.4.29}\eea
We obtained the last equality
in the conformal gauge, by combining the formulas (\ref{3.2.1})
and (\ref{3.2.6}) of the Appendix. Now we insert (\ref{3.4.29}) into
(\ref{3.4.28}) and in a straightforward way we arrive at the formula
(\ref{3.4.3}).

We may also check the validity of the formula (\ref{3.4.3}) for
a particular integrand. Indeed, let us compute the integral
\be\int D_{\mu,\phi}Q\exp{\big[{\ii\over 2\hbar}\I Q\p(\phi
\p)Q\big]}
={\rm det}^{-1/2}\big[-{\ii\over 2\hbar}
\e^{-\mu}{1\over \phi}\p(\phi \p)\big].\label{3.4.30}\ee
We have
(see Appendix Eqs.(\ref{3.2.9},\ref{3.2.10}))
\bea{\rm det}^{-1/2}\big[-{\ii\over 2\hbar}
\e^{-\mu}{1\over \phi}\p(\phi \p)\big]=
{\rm det}^{-1/2}\big[-{\ii\over 2\hbar}
{1\over \phi}\p(\phi \p)\big]\nn\\
\times \exp{\big\{-{\ii\over 48\pi}\I\{{1\over 2}\mu \p^2 \mu
+{3\over 2}\mu
[2\p^2\ln{\vert \phi \vert}+(\p\ln{\vert \phi \vert})^2] \} \big\} }.
\label{3.4.31}\eea
Because
\be {\rm det}^{-1/2}\big[-{\ii\over 2\hbar}
{1\over \phi}\p(\phi \p)\big]=
\int D_{0,\phi}Q\exp{\big[{\ii\over 2\hbar}\I Q\p(\phi \p)Q\big]},
\label{3.4.32}\ee
Eqs.(\ref{3.4.30}), (\ref{3.4.31}) and (\ref{3.4.32}) obviously
match with the formula (\ref{3.4.3}).

After this digression,
 we now compute the effective action for the case $N=24$. We use the
defining formula (\ref{3.1.7}) for the generating functional in the
conformal gauge and
insert the field dependences of the measures
(\ref{3.4.1}),(\ref{3.4.2})
and (\ref{3.4.3}) in it. We obtain

\bea W[J]=\int\! D_0 \mu~D_0 \phi ~ D_{0,\phi}Q^j
\exp{\big\{\!-{24\ii\over 32\pi}
\I\mu [2\p^2\ln{\vert \phi_c\vert} +(\p\ln{\vert
\phi_c\vert})^2]\big\}}\nn\\
\exp{\Big\{{\ii\over \hbar}
\I\big[{\phi\over 2\k}(-\p^2\mu -\k\p Q^j \p Q^j)
 -J_g\p^2\mu +J_{\phi}\phi +J_j Q_j\big]\Big\}}\nn\eea
\be\ \label{3.4.4}\ee
The integration over $Q^j$ is Gaussian and over $\mu$ gives the
$\delta$-function, therefore
\bea W[J]&=&
\int D_0 \phi~ \delta\big(\p^2\phi +2\k\p^2 J_g+{3\hbar\k\over 2\pi}
[2\p^2\ln{\vert \phi\vert} +(\p\ln{\vert \phi\vert})^2]\big)\nn\\
& &\times{\rm det}^{-12}\big[-{\ii\over 2\hbar}{1\over \phi}\p(\phi
\p)\big]
\exp{\big[{\ii\over\hbar}\I(J_{\phi}\phi -J_j{1\over 2\p(\phi
\p)}J_j)\big]}=
\nn\\&=&{\rm det}^{-12}\big[-{\ii\over 2\hbar}
{1\over \phi(J_g)}\p(\phi(J_g) \p)\big]\nn\\
& &\times\exp{\big[{\ii\over\hbar}\I(J_{\phi}
\phi(J_g) -J_j{1\over 2\p(\phi(J_g) \p)}J_j)\big]},\label{3.4.5}\eea
where the dependence of $\phi(J_g)$ on $J_g$ is dictated by the
$\delta$-function in (\ref{3.4.5}).
We stress that the formula (\ref{3.4.5}) gives the {\it exact}
generating
functional. Performing the Legendre transformation
(\ref{3.1.9}) and (\ref{3.1.10}) we obtain the {\it exact} effective
action
\bea \Gamma(\mu_c,\phi_c,Q^j_c)=\I{\phi_c\over 2\k}(-\p^2 \mu_c-\k
\p Q_c^j \p Q_c^j)\qquad\qquad \nn\\
-{3\over 4\pi}\I\mu_c[2\p^2\ln{\vert \phi_c\vert} +
(\p\ln{\vert \phi_c\vert})^2]
+12\ii\hbar ~{\rm lndet}\big[-{\ii\over 2\hbar}{1\over
\phi_c}\p(\phi_c \p)\big].\label{3.4.6}\eea
Comparing the result with Eq.(\ref{3.1.18}),
we conclude that for $N=24$ the semiclassical approximation is, in
fact,
exact.

\section{Quantum curvature singularities}
\subsection{The semiclassical field equations}

We obtain the semiclassical field equations by varying the one loop
effective action $\Gamma_1$ with respect to the classical fields
$\mu_c,\phi_c$ and $Q_c^j$. We have
\be -{1\over 2\k}\p^2\phi_c +\hbar{24\over 48\pi}\p^2\mu_c +
\hbar {\ii N\over 2}
{\delta\over \delta\mu_c}{\rm lndet}\big[-{\ii\over
2\hbar}\e^{-\mu_c}
{1\over \phi_c}\p(\phi_c \p)\big]=0,\label{3.3.1}\ee
\be \p(\phi_c \p Q_c^j)=0,\label{3.3.2}\ee
\be-\p^2\mu_c-\k\p Q_c^j \p Q_c^j+\hbar\ii\k N
 {\delta \over \delta\phi_c}{\rm lndet}\big[-{\ii\over 2\hbar}
\e^{-\mu_c}{1\over \phi_c}\p(\phi_c \p)\big]=0.\label{3.3.3}\ee
The solutions of the semiclassical equations have expansions
\be \mu_c=\mu_{c,0}+\hbar \mu_{c,1} + O(\hbar^2),\label{3.3.4}\ee
\be \phi_c=\phi_{c,0} +\hbar\phi_{c,1}+O(\hbar^2),\label{3.3.5}\ee
\be Q_c^j =Q_{c,0}^j +\hbar Q_{c,1}^j +O(\hbar^2).\label{3.3.6}\ee
Because we know just the first loop corrections to the effective
action,
the $O(\hbar^2)$-terms in the field expansions
(\ref{3.3.4}),(\ref{3.3.5})
and (\ref{3.3.6}) are irrelevant in the one loop approximation. Our
next
task will consist of the determination of $\mu_{c,1},\phi_{c,1}$ and
$Q_{c,1}^j$ from the semiclassical equations
(\ref{3.3.1}),(\ref{3.3.2})
and (\ref{3.3.3}), when $\mu_{c,0},\phi_{c,0}$ and $Q_{c,0}^j$ is
a given
classical solution. Since the ``lndet" terms in the one loop field
equations
are already of the order $O(\hbar)$, it is enough to compute the
functional derivatives of lndet at the {\it classical} solution
$\mu_{c,0}$
and $\phi_{c,0}(=t)$. The actual calculation requires the knowledge
of the
heat kernels of elliptic operators, some theory of the Hankel
transformations
and some integration of the Bessel functions. The details are
presented
in the Appendix here we list only the final result
\be{\delta\over \delta\mu_c}{\rm lndet}\big[-{\ii\over
2\hbar}\e^{-\mu_c}
{1\over \phi_c}\p(\phi_c \p)\big]={\ii\over 24\pi}\p^2\mu_c
+{\ii\over 16\pi}[2\p^2\ln{\vert \phi_c\vert} +(\p\ln{\vert
\phi_c\vert})^2
\big].\label{3.3.7}\ee
\bea {\delta \over \delta\phi_c(\xi)}{\rm lndet}\big[&-&{\ii\over
2\hbar}
\e^{-\mu_c}{1\over \phi_c}\p(\phi_c
\p)\big]\vert_{\phi_c=t}=\label{3.3.8}\\
&=& {\ii\over 8\pi t}
[\p^2\mu_c -\p(\mu_c \p\ln{\vert t \vert})\big]
-{\ii\over 8\pi t^3}
\ln{\big({\hbar t^2\over \ve\Omega}\big)},\nn\eea
where $\Omega$ is a finite constant and $\ve\to 0$ is the ultraviolet
cut-off. The following counter-term is needed for the cancelation of
the
UV divergence
\be {\cal L}_{count.}\sim \hbar\sq g^{\a\b}\phi^{-2}\p_{\a}\phi
\p_{\b}\phi.\label{3.3.9}\ee
This is the vertex of the new type, as it is usual in the
renormalization
of 2-d field theories. Its appropriate tuning replaces the product
$\ve\Omega$ in the semiclassical equations by some finite constant,
but it
cannot remove the logarithmical dependence on $t^2$.

Inserting the evaluated functional derivatives (\ref{3.3.7}) and
(\ref{3.3.8}) and the $\hbar$-expansion (\ref{3.3.4}),(\ref{3.3.5})
and (\ref{3.3.6}) of the fields into the semiclassical equations
(\ref{3.3.1}) , (\ref{3.3.2}) and (\ref{3.3.3}) we obtain the
equations
for $\mu_{c,1},\phi_{c,1}$ and $Q_{c,1}^j$
\be -{1\over 2\k}\p^2\phi_{c,1}+{24-N\over 48\pi}\p^2\mu_{c,o}
-{N\over
32\pi t^2}=0,\label{3.3.10}\ee
\be \p(\phi_{c,1}\p Q_{c,0}^j) +\p(t \p
Q_{c,1}^j)=0,\label{3.3.11}\ee
\be (-\p^2\mu_{c,1}-2\k \p Q_{c,1}^j \p Q_{c,0}^j)=
{N\k\over 8\pi}\big({1\over t}\p^2\mu_{c,0}+{1\over t^2}\p_t
\mu_{c,0}-
{1\over t^3}\mu_{c,0}-{1\over t^3}\ln{{\hbar t^2\over {\rm
const}}}\big).
\label{3.3.12}\ee
 This system of equations can be solved in a similar way as
 the classical system (\ref{2.2.3a}),(\ref{2.2.3b}) and
(\ref{2.2.3c}) was solved.
Indeed, because we know the Green function of the Minkowski
d'Alambertian, we find from Eq.(\ref{3.3.10}) the general form of
$\phi_{c,1}$, by adding arbitrary solution of the homogeneous
equation
to one particular solution of the full equation. Inserting
$\phi_{c,1}$
into (\ref{3.3.11}), we obtain the linear (inhomogeneous) Gowdy
equation for
$Q_{c,1}^j$. Since we know the eigenvalues and the eigenfunctions of
the Gowdy operator, we
know also its Green function and, eventaully, the general form of
$Q_{c,1}^j$.
Finally, putting $Q_{c,1}^j$ into (\ref{3.3.12}), we obtain the
linear
inhomogeneous d'Alambert equation for $\mu_{c,1}$, the general
solution
of which can be easily found. We conclude that our semiclassical
equations
(\ref{3.3.10}), (\ref{3.3.11}) and (\ref{3.3.12}) are exactly
solvable.
For our purposes, there is no need to write down the explicite
(and somewhat cumbersome) formulas. Instead of that we shall
concentrate
on the behaviour of the general solution near $t\sim 0$. We shall
show,
somewhat surprisingly, that even when we consider a regular classical
solution $\mu_{c,0},\phi_{c,0}$ and $Q_{c,0}^j$, the corresponding
solution
$\mu_{c,1},\phi_{c,1}$ and $Q_{c,1}^j$ possesses necessarily the
curvature
singularity at $t=0$. Such space-times are therefore classically
regular
but the quantum fluctuations induce the scalar curvature singularity,
proportional to $\hbar$. Hence, the quantum effects not only
do not smear the
classical curvature singularities, they even destabilize the regular
space-times! We present the corresponding quantitative analysis in
the next
subsection.

\subsection{The scalar curvature of the semiclassical space-times}

Let us study the behaviour of the scalar curvature $R$ near $t\sim 0$
for the space-times which solve the semiclassical field equations.
In this subsection we omit the index ``$c$" of the fields
$\mu_c,\phi_c$
and $Q^j_c$.
We choose such
classical metric
field $\mu_0$, that the classical space-time is nonsingular. From
Eqs.(\ref{2.2.12}) and (\ref{2.2.15}) for
$B^j=0$, it follows for $t\sim 0$
\be \mu_0\sim t^2 f(\s)+\cdots,\label{3.5.1}\ee
\be Q_0=g(\s)+t^2 h(\s)+\cdots,\label{3.5.2}\ee
where $f(\s),g(\s)$ and $h(\s)$ are functions, the concrete form of
which
is not relevant for our purposes and dots denote the subleading
terms,
also irrelevant for our analysis of the curvature singularities.
 Hence, from the Eq.(\ref{3.3.10}) we
find the behaviour of $\phi_1$ near $t\sim 0$
\be \phi_1 \sim -{N\k\over 16\pi}\ln{\vert t\vert}+{\rm
const}~f(\s)t^2
+F(U) +G(V)+\cdots. \label{3.5.3}\ee
The functions $F(U)$ and $G(V)$ cannot be specified from this
equation,
however, we may change our ``dilaton" gauge condition (\ref{2.2.8})
by the prescription
\be t+\hbar[F(U) +G(V)] \rightarrow t. \label{3.5.4}\ee
The one loop effective action (\ref{3.1.19}) is invariant under this
transformation, hence, the semiclassical equations
(\ref{3.3.1}),(\ref{3.3.2}) and (\ref{3.3.3}) remain unchanged.
Moreover,
the classical solution at which the functional derivatives of the
determinants are to be evaluated change just by the terms of order
$\hbar$.
This effect, of course, remain unseen in the one loop approximation,
because
the ``lndet" terms are already of the first order in
$\hbar$-expansion. Thus,
all subsequent analysis goes through and we can omit the
$F(U)+G(V)$ term and write without a loss
of generality
\be \phi_1 \sim -{N\k\over 16\pi}\ln{\vert t\vert}+{\rm
const}~t^2+\cdots
\label{3.5.5}\ee
Now we insert $\phi_1$ into Eq.(\ref{3.3.11}). We obtain
\be (-{\p^2\over \p t^2}-{1\over t}{\p\over \p t}+{\p^2\over
\p\s^2})Q_1
={H(\s)\over t}\ln{\vert t\vert}-{N\k h(\s)\over 8\pi t}+{\rm
bounded},
\label{3.5.6}\ee
where $H(\s)$ is some function of irrelevant shape. It is easy to
determine the behaviour of a particular solution of (\ref{3.5.6})
near
$t\sim 0$. It is given by
\be Q_{1,par}=-H(\s)t~\ln{\vert t\vert}+\big[2H(\s)+
{N\k h(\s)\over 8\pi}\big]t+\cdots
\label{3.5.7}\ee
A general solution of the $Q_1$ equation near $t\sim 0$ is then given
by
Eq.(\ref{3.5.7}) plus an arbitrary solution (\ref{2.2.12}) of the
homogeneous Gowdy equation. From (\ref{3.5.7}),(\ref{2.2.12}) and
(\ref{3.5.2}) then easily follows that
\be \p Q_1^j \p Q_0^j \sim C(\s)\ln{\vert t\vert}+{\rm
bounded}.\label{3.5.8}
\ee
The function $C(\s)$ vanishes, if the Neumann modes are absent in
the ``homogeneous" part of $Q_1$.
Inserting (\ref{3.5.8}) into the remaining semiclassical evolution
equation
(\ref{3.3.12}), we obtain
\be -\p^2\mu_1=2\k C(\s)\ln{\vert t\vert}-{N\k\over 8\pi}
\big[{f(\s)\over t}+{1\over t^3}\ln{{\hbar t^2\over {\rm
const}}}\big]
+{\rm bounded},\label{3.5.9}\ee
hence
\be \mu_1=-{N\k\over 16\pi t}\ln{{\hbar t^2\over{\rm const}}}+
\rho(U) +\nu(V) +{\rm bounded}.\label{3.5.10}\ee

In the classical case, the arbitrary integration functions
$\rho(U)$ and $\nu(V)$ are determined from the constraints
(\ref{2.2.9a}) and (\ref{2.2.9b}). In the semiclassical case they
have
to be determined from the given boundary conditions. The situation is
fully analogous to that occuring
 in the $CGHS$ model \cite{CGHS}, where the semiclassical
contribution to the constraints come from the Polyakov nonlocal
action
\be S_P=
+\hbar{24-N\over 96\pi}\I\sq R({1\over \sq}\p_{\a}
\sqrt{-g} g^{\a\b} \p_{\b})^{-1}R ,\label{3.5.11}\ee
We can make the generaly covariant action (\ref{3.5.11}) local at the
cost of introducing a new auxiliary field $Z$ (in a similar but not
identical
way as in \cite{Suss}). It reads
\be S_P=\hbar{24-N\over 48\pi}\I\sq\big[{1\over
2}g^{\a\b}\p_{\a}Z\p_{\b}Z
+RZ\big].\label{3.5.12}\ee
The contribution to the constraints are then obtained from $\delta
g^{uu}$
and $\delta g^{vv}$ variations of the action, in the conformal gauge.
We have
\be {1\over \sq}{\delta S_P\over \delta g^{uu}}=\hbar {24-N\over
48\pi}
({1\over 2}Z_u Z_u -Z_{uu}+\mu_u Z_u),\label{3.5.13a}\ee
\be {1\over \sq}{\delta S_P\over \delta g^{vv}}=\hbar {24-N\over
48\pi}
({1\over 2}Z_v Z_v -Z_{vv}+\mu_v Z_v),\label{3.5.13}\ee
We get rid of the auxiliary field $Z$ using the equation of motion
\be \p^2(\mu + Z)=0,\label{3.5.15}\ee
hence
\be \mu=-Z+\mu^+(u)+\mu^-(v)\label{3.5.106}\ee
and
\be{1\over\sq}{\delta S_P \over \delta g^{uu}}=\hbar {24-N\over
48\pi}
(\mu_{uu}-{1\over 2}\mu_u\mu_u)+T^+(u),\label{3.5.16}\ee
\be{1\over\sq}{\delta S_P \over \delta g^{vv}}=\hbar {24-N\over
48\pi}
(\mu_{vv}-{1\over 2}\mu_v\mu_v)+T^-(v),\label{3.5.17}\ee
The functions $T^+(u)$ and $T^-(v)$ are undetermined, because
$\mu^+(u)$
and $\mu^-(v)$ are not known.

In our present
model the Polyakov nonlocal action is the part of the one loop
semiclassical action (c.f.(\ref{3.1.19})). Therefore, the unknown
functions $\rho_(U)$ and $\nu(V)$ can be specified only by fixing the
boundary conditions. The $\delta g^{uu}$ and $\delta g^{vv}$
variations
of the remaining lndet part of the effective action of our model
cannot
influence this conclusion and we shall not consider them.

Finally we are ready to write down the scalar curvature of the
semiclassical
space-times. It reads (see Eq.(\ref{2.2.5}))
\bea
R(\hbar)&=&-\e^{-\mu(\hbar)}\p^2\mu(\hbar)=R(0)+\hbar(-R(0)\mu_1-
\e^{-\mu_0}\p^2\mu_1)+O(\hbar^2)=\nn\\
&=&R(0)+\hbar\big\{R(0){N\k\over 16\pi t}\ln{{\hbar t^2\over {\rm
const}}}
-R(0)(\rho(U)+\nu(V))\big\}\nn\\
& &+\hbar
\big\{-\e^{-\mu_0}2\k C(\s)\ln{\vert t\vert}+\e^{-\mu_0}{N\k\over
8\pi t}
(f(\s)+{1\over t^2}\ln{{\hbar t^2 \over \Omega}})\big\}\nn\\
& &+ {\rm bounded} +O(\hbar^2).\label{3.5.18}\eea
Clearly, whatever the functions $\rho(U)$ and $\nu(V)$
may be, the semiclassical
space-time is obviously singular. The singularity occurs at $t=0$ and
all
timelike observers will run into it - which is referred to as a
``thunderbolt" in Ref.\cite{HawkSte}.
 We arrived at the remarkable conclusion,
that while at the classical level there existed the nonsingular
spacetimes,
at the semiclassical level {\it all} space-times are necessarily
singular.
{}From Eq.(\ref{2.2.19}) and (\ref{3.5.18}) we also learn that the
singular
behaviour of classical and quantum curvature is different, hence no
cancelation of a classical curvature singularity due to quantum
effects
may occur\footnote{I am grateful to R.Jackiw for a comment on this
point.}.
If the classical spacetime is regular, then the formula
(\ref{3.5.18}) says,
that the corresponding semiclassical space-time
 is plagued by the curvature singularity proportional to $\hbar$.
Schematically
\be R= regular + \hbar ~ singular \cdots.\label{3.5.19}\ee
We conclude, that the quantum effects destabilize the classical
space-times
and lead to even more severe curvature singularities then the
classical
dynamics does. There remain only one possibility to avoid this
conclusion
in the framework of the present model, which may seem quite
unnatural,
however. It consists in introducing by hand
  into the effective action
 several $\it finite$ counter-terms of new
type, which would be fine-tuned so that to cancel the divergent terms
in (\ref{3.5.9}). But also keeping this possibility in mind we may
conclude
that the quantum instabilities in our model are generic.

\section{Conclusions and outlook}
We attempted to give a detailed description of the classical and
quantum
dynamics of the Jackiw-Teitelboim gravity with the cosmological
constant
replaced by the kinetic term of matter fields. We showed that the
classical
solutions of the model have natural physical interpretation, namely
they
describe the collisions of the wave packets of matter. For huge class
of such
solutions the corresponding classical space-times are topologically
trivial,
asymptotically flat
and free of curvature singularities. Then we computed the
semiclassical
effective action of the model; for the case $N=24$ we, in fact, got
the
exact expression. The effective field equations turned out to be
manageable
from the technical point of view. We have solved them and
 provided a simple analysis of the semiclassical
solutions near $t\sim 0$.The surprising result followed: the scalar
curvature acquires the quantum correction which is necessarily
singular.
Hence, the quantum fluctuations do not smear classical curvature
singularities, in fact they do just the opposite thing, they plague
the
regular classical space-times with quantum curvature singularities.
Because for $N=24$ we obtained the result starting from the exact
effective
action, our conclusion does not seem to be an artefact of the
semiclassical
approximation.

We believe that the model which we investigated is also interesting
from the
field theoretical point of view. At the classical level it is
completely
integrable and iteratively linear in the sense of subsection (2.2).
 This kind of ``linearity" played the
decisive role in the evaluation of the continual integral, in
a similar way as
it was
reported recently in the context of 2+1 Chern-Simmons theory
\cite{Gaw}.
The fact that for $N=24$ that computation gives the exact result
suggests an
 existence of a deeper {\it algebraic} structure in the
model\footnote
{This comment is due to K.Gaw\c edzki.}. Moreover, it turns out that
the model
possesses an unexpected and interesting {\it geometric} structure.
Indeed,
the action (\ref{2.1.4}) with the included matter field
 can be interpreted as the Jackiw- Teitelboim action (without
a cosmological
constant) in the non-commutative geometry of the ``two sheet"
manifolds
$Y\times Z_2$, where $Y$ is the 2-d space-time and $Z_2$ is the
internal
space containing just two points \cite{Fro1,Fro2}. The matter field
plays
the geometrical role of the distance between the two points in the
internal
space.

Our present model has also connections to the string theory on the
curved
backgrounds and to the exact 2-d conformal field theories. Indeed, in
the
conformal gauge, the classical action reads
\be S=-{1\over 2\k}\I(-\p\mu \p\phi+\k\phi \p Q^j \p
Q^j).\label{6.1}\ee
This is obviously an action of the non-linear $\sigma$-model
where $\mu,\phi$ and $Q^j$ are the coordinates of the target manifold
with the metric
\be \d s^2=-\d\mu \d\phi+\k\phi~ \d Q^j \d Q^j.\label{6.2}\ee
It is not difficult to see that the metric (\ref{6.2}) (it is written
in the
so called Rosen coordinates)
 describes a
{\it single} gravitational plane wave propagating on
$N+2$-dimensional
target! \cite{Garr,kliko} In other words, a {\it single}
gravitational wave in
$N+2$-dimensional target yields the $\sigma$-model action describing
collisions of the {\it two} gravitational waves in two dimensions.
Generalizing the work \cite{AmKl} Brooks
has shown that adding the target
dilaton background in the critical target dimension to such
a $\sigma$-model,
one obtains an exact
conformal field theory \cite{Broo}. It would be interesting to study
our present
model from this point of view. All the mentioned features of the
model look
quite promising for further investigations and we shall certainly
return to those
problems elsewhere.

\vskip .5 cm

I thank J.~Fr\"ohlich, K.~Gaw\c edzki and R.~Jackiw for enlightening
comments.

\section{Appendix}

In this appendix, we evaluate the functional derivatives of the
determinant in (\ref{3.1.18}), which were needed for obtaining the
explicit form of the semiclassical field equations. In evaluating the
Traces
we carefully keep in mind the definition of the scalar product
(\ref{3.1.4}).

Start with the derivative with respect to $\mu_c$.

\bea & & {\delta\over \delta \mu_c(\xi)}{\rm lndet}\big[-{\ii\over
2\hbar}
\e^{-\mu_c}{1\over \phi_c}\p(\phi_c \p)\big]=\nn\\
& &={\delta\over
\delta\mu_c(\xi)}(-1)\int_{\ve}^{\infty}{\d\tau\over\tau}
{\rm Tr}\exp{\Big\{-\tau\big[-{\ii\over 2\hbar}\e^{-\mu_c}{1\over
\phi_c}
\p(\phi_c \p)\big]\Big\}}=\nn\\& &
=\int_{\ve}^{\infty}\d\tau<\xi\vert{\ii\over
2\hbar}\e^{-\mu_c}{1\over \phi_c}
\p(\phi_c \p)\exp{\Big\{-\tau\big[-{\ii\over
2\hbar}\e^{-\mu_c}{1\over \phi_c}
\p(\phi_c \p)\big]\Big\}}\vert \xi>\nn\\& &
=-<\xi\vert \exp{\Big\{-\ve\big[-{\ii\over 2\hbar}\e^{-\mu_c}{1\over
\phi_c}
\p(\phi_c \p)\big]\Big\}}\vert \xi>.\label{3.2.1}\eea
We have used the heat-kernel regularization \cite{Haw,Dur} based on
the
following representation

\be\ln{x}=-\int_{\ve}^{\infty}{\d\tau\over\tau}\e^{-\tau x}+ (
{\rm an}~ x~ {\rm independent~ constant}) + O(\ve x).\label{3.2.2}\ee
The ``bras" and ``kets" in Eq.(\ref{3.2.1}) has to be understood in
the
standard sense.
The asymptotic expression for the heat kernel for small $\ve$ was
obtained for
an arbitrary elliptic operator in two dimensions \cite{Dur}. If M has
the form
\be M=-{1\over\sq}(\bigtriangledown_{\a}+B_{\a})\sq g^{\a\b}
(\bigtriangledown_{\b}+B_{\b})-B_0,\label{3.2.3}\ee
then\footnote{We put the sign + in front of $R$ (see also
Alvarez \cite{Alv} Eq.
(4.38)), because $R$ is given by
Eq.(\ref{2.2.5}).}
\be <\xi\vert \e^{-M\ve}\vert\xi>={1\over 4\pi\ve}\sq +{1\over
24\pi}R\sq
+{1\over 4\pi}B_0\sq +O(\ve).\label{3.2.4}\ee
Performing the Wick rotation to the Minkowski time, we can use
Eq.(\ref{3.2.4}) for evaluating the heat kernel (\ref{3.2.1}). In our
case
\be B_{\a}={1\over 2}\p_{\a}\ln{\vert \phi_c \vert},\label{rov}\ee
\be B_0=\e^{-\mu_c}\big[-{1\over 2}\p^2\ln{\vert
\phi_c\vert}-{1\over 4}
(\p \ln{\vert \phi_c \vert})^2\big],\label{3.2.5}\ee
hence
\be{\delta\over \delta\mu_c}{\rm lndet}\big[-{\ii\over
2\hbar}\e^{-\mu_c}
{1\over \phi_c}\p(\phi_c \p)\big]=\ii{1\over 24\pi}\p^2\mu_c
+{\ii\over 16\pi}[2\p^2\ln{\vert \phi_c\vert} +(\p\ln{\vert
\phi_c\vert})^2
\big].\label{3.2.6}\ee
Note that the functional derivative with respect to $\mu_c$ is the
local
expression. We also did not consider the first term in the r.h.s.~of
Eq.
(\ref{3.2.4}), which is eventually to be cancelled by the
two-dimensional
cosmological constant counter-term.

Next we compute the variation with respect to $\phi_c$. First of all
we
note that Eq.(\ref{3.2.6}) implies

\bea & & {\rm lndet}\big[-{\ii\over 2\hbar}\e^{-\mu_c}{1\over \phi_c}
\p(\phi_c \p)\big]=\nn\\ & &
\ii\I\big({1\over 48\pi}\mu_c \p^2 \mu_c +{1\over 16\pi}\mu_c
[2\p^2 \ln{\vert \phi_c \vert} +(\p\ln{\vert \phi_c \vert})^2]\big)
+F(\phi_c),
\label{3.2.9}\eea
where $F(\phi_c)$ is some $\mu_c$-independent functional. But
Eq.(\ref{3.2.9})
itself gives

\be {\rm lndet}\big[-{\ii\over 2\hbar}{1\over \phi_c}\p(\phi_c
\p)\big]=
F(\phi_c).\label{3.2.10}\ee
Hence,
\bea {\delta \over \delta\phi_c(\xi)}{\rm lndet}\big[-{\ii\over
2\hbar}
\e^{-\mu_c}{1\over \phi_c}\p(\phi_c \p)\big]= {\ii\over 8\pi \phi_c}
[\p^2\mu_c -\p(\mu_c \p\ln{\vert \phi_c \vert})\big]\nn\\
+{\delta\over\delta\phi_c(\xi)}{\rm lndet}\Big\{-{\ii\over
2\hbar}\big[\p^2
+(\p\ln{\vert \phi_c \vert})\p\big]\Big\}.\label{3.2.11}\eea
Now we have
\be\delta {\rm lndet}\Big\{-{\ii\over 2\hbar}\big[\p^2
+(\p\ln{\vert \phi_c
\vert})\p\big]\Big\}\vert_{\phi_c=t}=\label{3.2.12}\ee
$$-\delta\int_{\ve}^{\infty}{\d\tau\over\tau}{\rm Tr}
\exp{\Big\{-\tau\big[-{\ii\over 2\hbar}[\p^2+(\p\ln{\vert \phi_c
\vert})\p]
\big]\Big\}}\vert_{\phi_c =t}=$$
\bea\int_{\ve}^{\infty}\d\tau{\rm Tr}\Big\{-{\ii\over 2\hbar}\big(\p
{\delta \phi_c \over t}\big)\p\exp{\Big\{\tau\big[{1\over 2\hbar}
(\a-\ii)(\p_t^2 +{1\over t}\p_t){1\over
2\hbar}(\a+\ii)\p_{\s}^2\big]\Big\}}
\nn\eea
where $\a>0$ is a (small) ``Euclidean" cut-off, damping the
oscillatory
behaviour of the exponent. Now we wish to evaluate the last Trace in
Eq.(\ref{3.2.12}). Form the basis of the space of fields $Q$ as
follows

\be \Psi_{\pm,k,p}(t,\s)=\theta(\pm t)J_0(kt){1\over
\sqrt{2\pi}}\e^{-\ii p\s},
k>0,\label{3.2.13}\ee
where $\theta(t)$ is the usual step function. Using the theory of the
Hankel
transformations \cite{Gray}, it is easy to establish the relations of
orthogonality
\be \int_{R^2}\d t \d\s ~ t~ \Psi_{\pm k,p}^*(t,\s)\Psi_{\pm
k',p'}(t,\s)=
(-1)^{{\pm 1-1\over 2}}{1\over
k}\delta(k-k')\delta(p-p')\label{3.2.14a}\ee
\be \int_{R^2} \d t \d\s ~t~ \psi_{\pm k,p}^*(t,\s)\psi_{\mp
k',p'}(t,\s)=0.
\label{3.2.14b}\ee
and the relation of completeness
\be \int_0^{\infty}k \d k\int_{-\infty}^{\infty}\d p
\big(\Psi_{+
kp}^*(\xi)\Psi_{+kp}(\xi')-\Psi_{-kp}^*(\xi)\Psi_{-kp}(\xi')\big)
={1\over t}\delta (t-t')\delta(\s-\s').\label{3.2.15}\ee
(Note that in the r.h.s.~of Eq.(\ref{3.2.15}) stands the
$\delta$-function
$\delta(\xi,\xi ')$ with the respect to the inner product
(\ref{3.1.4})
for $\mu=0$ and $\phi=t$.)
Therefore, the Trace of an operator $O$ is given by
$${\rm Tr}O=\int_{R^2}\!\d t\d\s t \int_0^{\infty}\! k \d k
\int_{-\infty}^{\infty}\!\d p \big(\Psi_{+kp}^*(\xi)O\Psi_{+kp}(\xi)
-\Psi_{-kp}^*(\xi)O\Psi_{-kp}(\xi)\big)$$
\be\ \label{3.2.16}\ee
Using Eq.(\ref{3.2.16}), we can easily evaluate the Trace in
Eq.(\ref{3.2.12}). We have
\bea & & \delta {\rm lndet}\Big\{-{\ii\over 2\hbar}\big[\p^2+(\p
\ln{\vert \phi_c \vert})\p\big]\Big\}\vert_{\phi_c =t}=\nn\\
& & \int_{\ve}^{\infty}\!\d\tau \int_{R^2}\d t \d\s
t \int_0^{\infty}\!
k \d k \int_{-\infty}^{\infty}\!\d p\exp{\Big\{\!-\tau\big[{1\over
2\hbar}
(\a-\ii)k^2 +{1\over 2\hbar}(\a+\ii)p^2\big]\Big\}}\nn\\
& &\big(\Psi_{+kp}^*(\xi)(-{\ii\over 2\hbar}(\p{\delta\phi_c\over
t})\p)
\Psi_{+kp}(\xi)-
\big(\Psi_{-kp}^*(\xi)(-{\ii\over 2\hbar}(\p{\delta\phi_c\over t})\p)
\Psi_{-kp}(\xi)\big)=\nn\\& &
=\int_{\ve}^{\infty}\d\tau \int_{R^2}\d t \d\s t \int_0^{\infty}k \d
k
\int_{-\infty}^{\infty}\d p\exp{\Big\{-{\tau\over 2\hbar}
\big[(\a-\ii)k^2 +(\a+\ii)p^2\big]\Big\}}\nn\\& &
{\ii\over 2\hbar }(\p_t {\delta \phi_c \over t}){1\over 2}\p_t
\big(\Psi_{+kp}^*(\xi)\Psi_{+kp}(\xi)-\Psi_{-kp}^*(\xi)\Psi_{-kp}
(\xi)\big).
\label{3.2.17}\eea
Now the formula (see \cite{Gray})
\be \int_0^{\infty}r\d r \e^{-\rho^2 r^2} J_0(\lambda r)J_0(\mu r)=
{1\over 2\rho^2}\e^{-{(\lambda^2 +\mu^2)\over
4\rho^2}}J_0({\ii\lambda\mu
\over 2\rho^2})\label{3.2.18}\ee
and the standard Gaussian integration explicitly give the integrals
over
$k$ and $p$. We obtain (for $t\neq0$)
\bea \delta {\rm lndet}\Big\{-{\ii\over 2\hbar}\big[\p^2+(\p\ln{\vert
\phi_c \vert})\p\big]\Big\}\vert_{\phi_c =t}=
\int_{\ve}^{\infty}\d\tau \int_{R^2}\d t \d\s t {\ii\over 4\hbar}
\p_t({\delta \phi_c\over t})\nn\\\sqrt{\hbar\over (\a+\ii)2\tau\pi}
\p_t\Big[{\hbar\over (\a-\ii)\tau}\e^{-{\hbar t^2\over (\a-\ii)\tau}}
J_0\big({\ii\hbar t^2\over \tau (\a-\ii)}\big)
\big(\theta^2(t)-\theta^2(-t)\big)\Big].\label{3.2.18b}\eea
We can rewrite Eq.(\ref{3.2.18b}) as follows
\bea & & {\delta \over \delta \phi_c(t,\s)}{\rm lndet}
\Big\{-{\ii\over 2\hbar}\big[\p^2+(\p\ln{\vert \phi_c
\vert})\p\big]\Big\}
\vert_{\phi_c = t}=\nn\\
& & = -[\theta(t)-\theta(-t)]\int_0^{\hbar\over\ve}\d \rho
\rho^{-1/2}
{\ii\over 4\sqrt{2\pi(\a-\ii)}}{1\over t}\p_t\Big[t\p_t
\big(\e^{-{\rho t^2\over \a-\ii}}J_0\big({\ii\rho t^2\over
\a-\ii}\big)\big)
\Big]\nn\\
& &=-[\theta(t)-\theta(-t)]{\ii\over 4\sqrt{2\pi(\a-\ii)}}{1\over
t}\p_t t
\p_t ({1\over \vert t \vert}\int_0^{{\hbar t^2\over \ve}}\d\rho
\rho^{-1/2} \e^{-{\rho\over \a-\ii}}J_0({\ii\rho\over
\a-\ii})\big)\nn\\
& &=-{\ii\over 4\sqrt{2\pi(\a-\ii)}}{1\over t}\p_t t \p_t {1\over t}
\int_0^{\hbar t^2/\ve}\d\rho \rho^{-1/2} \e^{-{\rho\over \a-\ii}}
J_0({\ii\rho\over \a-\ii}).\label{3.2.19}\eea
Now we can decompose the integral over $\rho$
\bea & & \int_0^{\hbar t^2/\ve}{\d\rho \over \sqrt{\rho}}
\e^{-{\rho\over
\a-\ii}}J_0({\ii\rho\over \a-\ii})=\label{3.2.20}\\& &
{\rm const} +\int_1^{\hbar t^2 /\ve}{\d\rho\over
\sqrt{\rho}}\e^{-{\rho
\over \a-\ii}}\bigl(J_0({\ii\rho\over \a-\ii})-{\a-\ii\over
\sqrt{2\pi\rho}}
\e^{\rho\over \a-\ii}\bigl)+\int_1^{\hbar t^2/\ve}{\d\rho\over\rho}
\sqrt{{\a-\ii \over 2\pi}}.\nn\eea
{}From the asymptotic behaviour of the Bessel functions \cite{Gray},
we conclude that the first integral in the r.h.s.~of
Eq.(\ref{3.2.20})
is convergent for $\ve\to 0$. Hence
\bea & & {\delta \over \delta \phi_c(t,\s)}{\rm lndet}
\Big\{-{\ii\over 2\hbar}\big[\p^2+(\p\ln{\vert \phi_c
\vert})\p\big]\Big\}
\vert_{\phi_c = t}=\nn\\& &
-{\ii\over 8\pi} {1\over t}~\p_t ~t ~\p_t {1\over t}\ln{\big({\hbar
t^2
\over \ve~ {\rm const}}\big)}=-{\ii\over 8\pi t^3}
\ln{\big({\hbar t^2\over \ve\Omega}\big)},\label{3.2.21}\eea
where $\Omega$ is a finite constant.

\end{document}